\documentstyle[aps,pre,preprint]{revtex}

\begin{document}

\newcommand {\good } {\mbox{${\rm good}$}  }
\newcommand {\all  } {\mbox{${\rm all }$}  }
\newcommand {\tgap } {{\rm true gap}       }
\newcommand {\xgap } {{\rm mxmgap}         }
\newcommand {\vol  } {{\rm vol   }         }
\newcommand {\Om   } {\Omega               }
\newcommand {\Ga   } {\Gamma               }
\newcommand {\ga   } {\gamma               }
\newcommand {\s    } {\sigma               }
\newcommand {\bsi  } {{\mbox{\boldmath$\s$}}}
\newcommand {\lc   } {\left\{              }
\newcommand {\rc   } {\right\}             }
\newcommand {\beq  } {\begin{equation}     }
\newcommand {\eeq  } {\end{equation}       }
\newcommand {\bea  } {\begin{eqnarray}     }
\newcommand {\eea  } {\end{eqnarray}       }
\newcommand {\nn   } {\nonumber            }
\newcommand {\hsp  } {\hspace              }
\newcommand {\hsc  } {\hspace{1cm}         }
\newcommand {\hsm  } {\hspace{15mm}        }
\newcommand {\hsh  } {\hspace{5mm}         }
\newcommand {\vsp  } {\vspace              }
\newcommand {\nl   } {\newline             }
\newcommand {\ev   } {\equiv               }
\newcommand {\vp   } {{\vec{p}}            }
\newcommand {\ov   } {\over                }
\newcommand {\ha   } {\mbox{${1\ov2}$}     }
\newcommand {\e    } {\!+\!                }
\newcommand {\m    } {\!-\!                }
\newcommand {\lb   } {\label               }
\newcommand {\lh   } {\left(               }
\newcommand {\rh   } {\right)              }
\newcommand {\noi  } {\noindent            }
\newcommand {\Og   } {\Om^{\rm good}_\Ga   }
\newcommand {\Oa   } {\Om^{\rm  all}_\Ga   }
\newcommand {\Zh   } {\langle Z\rangle_{\rm harm}}
\newcommand {\Eh   } {\langle E\rangle     }

\title{Determination of Interaction Potentials of Amino Acids from Native 
Protein Structures: Tests on Simple Lattice Models.}
\author{\em Jort van Mourik $^1$, Cecilia Clementi $^1$, Amos Maritan $^1$,}
\author{\em Flavio Seno $^2$, Jayanth R. Banavar $^3$}
\author{}
\address{$^1$ International School for Advanced Studies (SISSA) 
and Istituto Nazionale di Fisica della Materia,
Via Beirut 2-4, 34014 Trieste, Italy}
\address{$^2$ Dipartimento di Fisica G. Galilei Universit{\`a} di Padova
and Istituto Nazionale di Fisica della Materia,
Via Marzolo 8, 35131 Padova, Italy}
\address{$^3$ Department of Physics and Center for Material Physics
104 Davey Laboratary, The Pennsylvania State University,
University Park, PA 16802 -USA}

\date{\today}
\maketitle
\centerline{PACS numbers: 87.15.By, 87.15.Da, 87.10.+e, 05.20.-y}
\begin{abstract}
\noi We propose a novel method for the determination of the effective 
interaction potential between the amino acids of a protein. 
The strategy is based on the combination of a new optimization procedure
and a geometrical argument, which also uncovers the shortcomings of any
optimization scheme. The strategy can be applied on any data set of 
native structures such as those available from the Protein Data Bank (PDB).
In this work, however, we explain and test our approach on simple lattice 
models, where the true interactions are known a priori. Excellent agreement
is obtained between the extracted and the true potentials even for modest 
numbers of protein structures in the PDB. Comparisons with other methods 
are also discussed.  
\end{abstract}

\section{Introduction                                          \lb{intro}}

The prediction of the three dimensional structures of the native state of 
proteins from the knowledge of their sequences of amino acids can only be 
achieved if the interaction potentials among various parts of the peptide chain
in the presence of solvent molecules are known to some extent.
Indeed, the native states of many globular proteins correspond to the
conformations which are global minima of the free energy \cite{An}.
Thus the knowledge of the energy of a sequence in a given 
conformation would be an important step in the complete solution 
of this formidable problem, 
and also of the inverse one, i.e. the design of a sequence of amino acids 
that rapidly folds into a desired conformation.
\nl
A rigorous approach from ``first principles'', taking into account the 
quantum mechanics  of the huge number of atoms constituting the protein is 
not practical and beyond actual possibilities.
\nl
An alternative approach consists of introducing a coarse-grained 
description mainly based on lattice models where the peptide chain is 
a self avoiding walk whose nodes represent extremely 
simplified amino acids. Models of this type have been widely used in the 
recent literature for various goals, ranging from folding dynamics to 
thermodynamic properties of folded states of proteins, see e.g. 
\cite{SS,DB,GO}\ .
\nl 
One of the main difficulties with such simplified representations of the
protein chain is the fact that an effective interaction Hamiltonian has to
be used, which captures the essential features of the specific properties 
that one wishes to describe.  For example, it is commonly believed that the
native states of protein sequences ought to correspond to pronounced minima
in conformation space \cite{SS}. In the most commonly used model 
Hamiltonian,  ``effective'' two-body forces between neighboring (in space
but not in sequence) amino
acids are the only interactions that are considered \cite{TS}-\cite{MS}. These  ``effective'' forces also take solvent 
induced interactions into account.
\nl
Traditionally, the potential energies of the interactions have been derived
from pairing frequencies of amino acids observed in the native structures 
contained in the PDB \cite{TS,MJ,XX,TD}. The method, known as the quasi-chemical
approximation, is widely used and relatively easy to implement in such a 
difficult context. In important recent work, Thomas and Dill \cite{TD} 
have rigorously tested the underlying assumptions and approximations of 
the quasichemical method. Employing a lattice model with an a priori 
assigned interaction potential, one is able to construct a PDB identifying 
proteins as amino acid sequences having a unique ground state conformation 
(native state) among all possible conformations (this is accomplished by
exhaustive exact enumeration for sufficiently small values of
the protein chain length and/or the number of 
amino acid classes). Applying the quasichemical method to 
several of these exact cases, Thomas and Dill \cite{TD} demonstrated 
the inadequecies of the method and identified its possible weak points. 
Indeed, the interaction parameters could, in the worst cases, be off by as 
much as a factor of two and the native states of protein sequences of the 
PDB could be correctly identified onaverage in 84\% of 
the cases which is poor, for the simple model employed.
\nl
Recently, we have proposed \cite{SM} a new optimization method for the 
determination of effective potentials --  the derived potentials in the model considered
by Thomas and Dill \cite{TD} are better than the ones obtained by the 
quasichemical method, but still do not match the true potentials. 
Nevertheless, these 
derived potentials allow 100\% success in the prediction of the native 
structures!
In this work we explain why this is so, and show that with the bare
information contained in the native structures, no unique value of the
candidate potentials can be given. Starting from a set of "good"
sequences, i.e. sequences which have a unique ground state with the true
potentials, their corresponding native structures, and a set of
alternative structures, we can isolate a volume (cell) in the space of
potentials. All points in this cell are equivalent to the true potential
as far as the only requirement is that each good sequence has to
recognize (i.e. has as the unique ground state) its native structure.
The volume of the cell decreases as the protein chain length increases.
In order to identify the most likely point around which the cell shrinks,
we have to come up with some criterion. Here we propose a new implementation
which leaves
the success of good sequences in finding their own native states 
unaffected, and improves considerably the estimate of the extracted
potentials.\nl
As already stressed in \cite{SM}, our method has its root in the original
proposal by Crippen \cite{Cr} but differs substantially in the 
implementation \cite{MC}. Our method is general, it can be implemented at
any desired temperature $T$ (lower than the minimum folding transition 
temperature of the good sequences we are considering), and it does not have
any adjustable parameter. The method is  explained in Sec. 2\ , whereas 
Sec. 3 contains the results for a 2- and 4-class amino acid problem. Since 
our method is applicable in any spatial dimensionality,
we have restricted ourselves to various checks in a two-dimensional square 
lattice  with chains up to length 16 restricted to lie within a $5\times5$ square.  A 
comparison of our results with those obtained using a recently proposed 
method  of Mirny and Shakhnovich (\cite{MS}) is made in Sec. 4\ . 

\section{The Model                                            \lb{model}}

We consider a set of $N_s$ sequences $\Om_\s=\{\bsi_s\}_{s=1,..,N_s}$
each comprised  of 
$N$ amino acids $\bsi_s=\{\s^s_i\}_{i=1,..,N}$. 
Each sequence $\bsi_s$ is postulated to have a unique native state (assumed 
to be the ground state and denoted by the superscript $n$) in a spatial conformation $\Ga^n_s$ that is known 
experimentally or otherwise. The corresponding set of native conformations 
is denoted by $\Om_\Ga=\{\Ga^n_s\}_{s=1,..,N_s}$. 
\nl
We assume that for a given number of amino acid types $N_a$, the effective
interaction potentials can be written in the form of a symmetric 
interaction matrix $P_{\mu\nu}$, $\mu,\nu=1,..,N_a$\, and that
similarly for each combination of a sequence and a conformation, a 
symmetric contact matrix $C(\Ga,\bsi)_{\mu\nu}$, $\mu,\nu=1,..,N_a$ is 
defined, giving the (effective) number of contacts between the different 
types of amino acids. The energy of a sequence $\bsi$ in conformation $\Ga$
is thus given by

\beq
E(\Ga,\bsi)=\sum^{N_a}_{\mu\leq\nu=1} P_{\mu\nu}\; C(\Ga,\bsi)_{\mu\nu}
\ . \lb{E}
\eeq

\noi There are a gigantic number of spatial conformations a sequence $\bsi$
can take, which we label as $\Ga_i(\bsi)$, and $\Ga^n(\bsi)\ev\Ga_0(\bsi)$ 
is the experimentally determined native state structure. At a temperature 
$T$, the probability that the sequence is in one of these conformations is 
simply given by 

\beq
P_i(\bsi) = \exp\left[-(E(\Ga_i(\bsi),\bsi)-F(\bsi))/T\right]\ , \lb{Pi}
\eeq

\noindent
where the Boltzmann constant $k_B$ is defined equal to 1, and $F(\bsi)$ is 
the free energy, defined as

\beq
F(\bsi) = -T \log\lh\sum_i\exp[-E(\Ga_i(\bsi),\bsi)/T]\rh\ .
\eeq

\noi Because the experimentally observed structure of the sequence is 
in the conformation $\Ga^n(\bsi)$, the value of $P_0(\bsi)$ must be large 
$(>\ha)$ at temperatures below the folding transition temperature. Indeed, $P_0(\bsi)$
should be equal to 1 at zero temperature, if the ground state is
non-degenerate. In recent work, Crippen \cite{C2}
has suggested that even with the knowledge of the exact contact potential 
from which the folding sequences and their unique native conformations are 
determined, one may not be able to correctly select which sequences fold to
a desired target structure. The resolution \cite{DK} of this puzzle is that 
the right ``score" to be maximized in the inverse folding problem is 
(\ref{Pi}), i.e. $E(\Ga(\bsi),\bsi)-F(\bsi)$ has to be minimized, and not 
just the energy $E(\Ga(\bsi),\bsi)$. 
A key feature of this score is that $F(\bsi)$, the free energy, does
not depend on the specific target structure, but only on the sequence being
considered. Thus, the determination of the exact contact potential is a 
valuable first step for an attack on the protein design problem, even 
though the currently used score needs to be modified.
\nl
In what follows, we describe a zero temperature version (which is 
appropriate in most instances) of such a procedure to extract the exact
potentials. Furthermore, we restrict ourselves to models where each 
conformation is a self-avoiding walk whose elementary steps join nearest 
neighbor sites of a $d$-dimensional hypercubic lattice ($d=2$ in the
present applications). Amino acids are 
placed at the nodes of the visted sites, and contacts are defined between
amino acids in neighboring sites but not next to each other along the sequence. 

\subsection{The Method                                       \lb{method}}

\noi Instead of starting immediately with a cost function that has to be 
minimized, we concentrate for a moment on the space spanned by the 
interaction potentials.\nl 
Since all energies scale linearly with the amplitude of the interaction
potentials, we have to keep e.g. the first parameter fixed 
($P_{11}\to P_0$) to set a scale. Relabelling the remaining parameters 
$P_{\mu\nu}\to \vp\ev\{p_i\}_{i=1,..,N_p}$ ({\footnotesize $N_p\ev\ha 
N_a(N_a\e1)\m1$}), and renumbering the contacts accordingly, 
we can rewrite (\ref{E}) as

\beq
E(\Ga,\bsi)=\sum_{i=1}^{N_p} p_i c_i(\Ga,\bsi)+P_0 c_0(\Ga_,\bsi)\ev
\vec{p}\cdot\vec{c}(\Ga,\bsi)+P_0 c_0(\Ga_,\bsi)\ .
\eeq

\noi  The fact 
that a sequence has a lower energy in its native conformation than in any 
alternative conformation, provides a linear inequality (or hyperplane) in 
the parameter space separating the space into  allowed and  forbidden 
halfspaces.  We define:

\beq
I_i(\Ga^{\rm alt},\Ga^n_s,\bsi_s)\ev c_i(\Ga^{\rm alt},\bsi_s)-
c_i(\Ga^n_s,\bsi_s)\ .
\eeq

\noi The allowed points in parameter space have to satisfy the linear 
inequality

\beq
\sum_{i=1}^{N_p} p_i I_i+P_0 I_0\ev\vec{p}\cdot\vec{I}+P_0 I_0 > 0\ .
\eeq

\noi Repeating this operation for all the sequences in $\Om_\s$ and 
for all the alternative conformations, and retaining only the allowed part 
of the parameter space that satisfies all the inequalities, we obtain a 
convex {\em cell} around the target parameters. This cell contains all the 
points that yield the correct native conformation as the unique ground state 
for each of  the sequences in $\Om_\s$. In the test model, we have generated 
the set $\Om_\s$ using the energy function (\ref{E}), and therefore, the 
existence of the cell is guaranteed and the problem is well posed. For real
proteins the form of the energy function is an ansatz that is tested by the
(non)existence of a finite cell.\nl
Each inequality corresponds to a hyperplane in parameter space separating
allowed and  forbidden half-spaces. The 
orientation of the hyperplane 
is given by $\vec{I}$, the offset from the origin by 
$P_0 I_0$. The distance of any point $\vec{p}$ in parameter space to this
hyperplane, is related to the energy gap between the two configurations
leading to this inequality (at the value of parameters given by $\vec{p}$)
by the following linear equation: 

\beq
{\rm d}(\vp,I)={\left| \vec{p}\cdot\vec{I}+P_0 I_0\right|\ov\sqrt{\sum_i 
I^2_i}}={{\rm gap}(\vp,I)\ov\sqrt{\sum_i I^2_i}}\ . \lb{d}
\eeq

\noi Using all the information in the data set, the cell is maximally 
reduced.  A selection procedure is needed in order to isolate an optimal 
point within the cell.  Instead of the cost functions used in \cite{SM}, 
the optimal interactions are chosen such that the smallest gap among all 
the sequences in the training set is as large as possible.  The cost 
function ($F_{\rm gap}(P)$) is hence taken to be minus the smallest gap, 
i.e.

\bea
F_{\rm gap}(P)&=&-\min_{\bsi_s\in\Om_\s}\min_{\Ga\neq
\Ga^n_s}\;\lc E(\Ga,\bsi_s)-E(\Ga^n_s,\bsi_s)\rc  \nn \\
&=&-\lh\vec{p}\cdot\vec{I}(\Ga^*,\Ga^n_{s^*},\bsi_{s^*})+
P_0I_0(\Ga^*,\Ga^n_{s^*},\bsi_{s^*})\rh\ ,
\lb{F}
\eea

\noi where $\bsi_{s^*}$ and $\Ga^*$ are the sequence and the alternative 
conformation respectively that yield the minimum gap. To reiterate, the 
interaction potentials are chosen in such a way that the maximum minimum 
(mxm-) gap is obtained. For similar ideas see also \cite{XY}.
\nl 
This cost function has two major advantages over previous attempts.  First, it automatically ensures that all sequences have 
their unique groundstates in the correct structures. In fact, a negative 
mxm-gap would imply that the a priori assumption of the form of the energy 
function (\ref{E}) is incompatible with the data in the training set.
\nl
Second, it does not suffer from an unphysical bias due to statistical
fluctuations that were present in the cost functions proposed in \cite{SM}. 
These cost functions  not only make use of all inequalities, but also of 
the number of occurences of each inequality over the training set. 
Therefore, it may happen that inequalities that occur more frequently, 
push the optimal parameters away from their true values. 
One may expect all sequences in the training set to satisfy the minimimum 
conditions that make them good folders, which implies that each inequality
is equally important, irrespective of the number of times it occurs. 
In realistic cases it may be important to rescale the energy gap associated 
with a given sequence with respect to its ground state energy. Because, in 
this work, we have sequences of the same length in a given training set, 
all ground state energies are roughly the same, and rescaling of the gap is
not necessary.
\nl
Because the width of the energy gap is not equal to the distance in parameter space
(\ref{d}), inequalities that do not contribute to the boundaries of the 
cell may still influence the mxm-gap. Nevertheless, for all the cases that 
we have 
encountered, using only the inequalities contributing to the cell, we 
obtain exactly the same optimal parameters as the ones derived by using all
inequalities. This facilitates our maximization procedure. The sequences 
have to be put on each alternative conformation only once before the 
optimization procedure, and we retain only those inequalities that 
contribute to the cell. Then we start our optimization procedure with only 
those few inequalities. Given the inequalities with respect to which to 
optimize, the design of an optimization procedure is straightforward, 
because the gradient of each inequality is given by $\vec{I}$. Therefore, 
each step in the optimization procedure consists of the following 
replacement

\bea
\vec{p}&\to&\vec{p}+\ga \vec{I}(\Ga^*,\Ga^n_{s^*},\bsi_{s^*}) \ ,\ {\rm or}
\lb{op1} \\ 
\vec{p}&\to&\vec{p}+\ga \vec{I}_{\rm min}                  \ ,\lb{op2} 
\eea

\noi where $\ga$ is a parameter that can be tuned to obtain fast 
convergence. Form (\ref{op1}) of the optimization algorithm is used when the
equalities have to be recalculated putting each good sequence on the 
alternative conformations, while form (\ref{op2}) is used when the set of 
important inequalities is already known. In that case $I_{\rm min}$ is the 
inequality from this set that yields the minimum gap.

\section{Results                                            \lb{res}}

\noi The method has been tested on models with an increasing number of 
interaction parameters to check the dependence on the dimensionality 
($N_p$) of the parameter space. The test has been done on the normal 
H-P model \cite{LD}, $N_a=2$, with nearest neighbor (nn) interactions, i.e. 
with 2 free parameters ($N_p=2$), and some variations like considering next
to nearest neighbor interactions, to check the robustness of the method.
\nl
Furthermore, the method has been applied to models with 4 types of amino 
acids ($N_p=9$) and nn interactions. For the latter we have also studied 
the dependence of the quality of the obtained parameters on the size of 
the PDB and of the set of alternative structures.
\nl\nl  
Although still feasible up to parameter numbers as high as $N_p=9$, 
increasing the number of interaction parameters and thus the dimension of 
the cells, reveals the tendency that the advantage of putting the good 
sequences on the alternative structures only once, will be annihilated by 
having to calculate too many cornerpoints.
\nl 
With increasing dimension, the number of inequalities contributing to the 
cell grows linearly, while the number of cornerpoints of the cell, however, 
grows exponentially. Therefore, one may have to opt for a hybrid method.
The first step consists of a rough optimization recalculating the 
inequalities at each update (\ref{op1}). Once a point in 
parameter space satisfying all constraints (i.e. in the cell) has been
obtained, we select 
and save the inequalites which are at a distance less than some tolerance
parameter 
from this point. The number of such inequalities is relatively low and it grows linearly with the 
number of parameters. Then, we fully optimize only with respect to these 
inequalities (\ref{op2}). An implementation of this method on the model with 
$N_p=9$ for different choices of parameters shows that we roughly need 
20-30 updates for the first step. Then, we need to typically
save of the order of 100
inequalities to perform the second step.\nl
This hybrid method is very efficient because it uses the insights in the 
geometry of the cells in parameter space and avoids unnecessary time loss 
due to exactly calculating the cell.

\subsection{Results for the H-P model}                  \lb{reshp}

\noi In order to be able to compare our results with those of previous 
work, the first model, that we study extensively, is the H-P model 
introduced by Dill and co-workers \cite{LD}, which has 2 types of amino 
acids. A contact is defined to be 1 for a nearest neighbour contact, as long
as the two amino acids are next to each other along the sequence,
and 0 otherwise. To fix the energy scale, we 
have choosen to fix the parameter ($E_{HH}\ev P_0$). Hence, we are left 
with two independent parameters ($E_{PP}\ev p_1$ and $E_{HP}\ev p_2$) and a
2 dimensional parameter space, which allows us to clarify our reasoning 
with instructive pictures. We have considered three types of target 
interaction parameters: $(E_{HH},E_{PP},E_{HP})=(-1,0,0),\ (-1,-1/\sqrt{2}
\!\simeq\!-0.707,0),\ (-2,-2,-1)$, and seven distinct groups of amino acid 
chains each of length: $N=10,..,16$.
\nl\nl
In order to reduce the necessary computer time, we have taken all 
semi-compact 2 dimensional conformations on a square lattice.  By 
semi-compact, we mean that we restrict the conformations to a box of size 
$5\times5$ (Tests with all conformations of a certain length on a square 
lattice show that the results are unaltered).
\nl
As alternative conformations, we have considered both the set of good 
conformations $\Og$ (having at least one sequence that has its unique
native state
in it), and the set of all conformations $\Oa$ (obtained by complete 
enumeration and also used to generate the good sequences). Although good 
results can be obtained considering only $\Og$, it is not excluded (and is 
indeed observed) that extra information can be gained (in the sense of new 
inequalities, further reducing the cell) by also considering new 
conformations from $\Oa$. However, when the cells are closed, in all cases
we obtain exactly the same set of optimized parameters. Since some 
inequalities are very rare, it is difficult to say how many alternative 
conformations are needed to maximally reduce the cell. Unfortunately, so 
far we have not found a criterion to determine  beforehand whether a certain 
sequence and a given
alternative conformation gives rise to a ``tight'' inequality 
(contributing to the boundaries of the final cell) or not. Therefore, 
although the obtained parameters are relatively stable to changes in the 
shape of the cell, the best strategy seems to be to use as much information as is 
available, or as is numerically feasible. It cannot be excluded that 
regenerating all the good sequences with the newly obtained parameters, 
would add some new ``good'' sequences to the set.
\nl

\renewcommand{\theequation}{3.1.\alph{equation}}
\setcounter{equation}{0} 

\beq
\begin{tabular}
{||l | l       | l         | l        | l       | l        | l          ||}
\hline
\hline
{\rm target}& 0.0& 0.0       &          &         &          &          \\ 
\hline
 $N$&$E_{PP}\hsc$&$E_{HP}\hsm$&$\tgap\;$&$\xgap$ &$\vol(\Oa)$&$\vol(\Og)$\\
\hline
 10 & /        & /         & 1.0      & /       & 0.0$^o$  & 2.034444$^o$\\
 11 & 0.0      & 0.0       & 1.0      & 1.0     & 1.062500 & 0.643501$^o$\\
 12 & 0.0      & 0.0       & 1.0      & 1.0     & 0.916667 & 0.643501$^o$\\
 13 & 0.0      & 0.0       & 1.0      & 1.0     & 0.625000 & 0.0$^o$     \\
 14 & 0.0      & 0.0       & 1.0      & 1.0     & 0.444444 & 0.0$^o$     \\
 15 & 0.0      & 0.0       & 1.0      & 1.0     & 0.272817 & 1.203704    \\
 16 & 0.0      & 0.0       & 1.0      & 1.0     & 0.252315 & 0.611111    \\
\hline
\hline
\end{tabular} 
\lb{tab31a} 
\eeq

\beq
\begin{tabular}
{||l | l       | l         | l        | l        | l       | l          ||}
\hline
\hline
{\rm target}&-0.707107&0.0 &         &          &          &          \\ 
\hline
 $N$&$E_{PP}\hsc$&$E_{HP}\hsm$&$\tgap\;$&$\xgap$&$\vol(\Oa)$&$\vol(\Og)$ \\
\hline
 10 & /        & /         & 0.12132 & /        & 0.0$^o$  &0.321751$^o$\\ 
 11 & -5/7     & 0.0-0.10  & 0.12132 & 0.142856 & 0.034722 & 0.034722   \\
 12 & -5/7     & 0.0-0.04  & 0.12132 & 0.142856 & 0.017361 & 0.029514   \\
 13 & -5/7     & 0.0       & 0.12132 & 0.142856 & 0.030556 & 0.030556   \\
 14 & -5/7     & 0.0       & 0.12132 & 0.142856 & 0.015129 & 0.019097   \\
 15 & -5/7     & 0.0       & 0.12132 & 0.142856 & 0.007955 & 0.007955   \\
 16 & -5/7     & 0.0       & 0.12132 & 0.142856 & 0.007955 & 0.007955   \\
\hline
\hline
\end{tabular}
\lb{tab31b} 
\eeq

\beq
\begin{tabular}
{||l | l       | l         | l        | l       | l        | l          ||}
\hline
\hline
{\rm target}& -2.0& -1.0     &          &         &          &          \\
\hline
 $N$&$E_{PP}\hsc$&$E_{HP}\hsm$&$\tgap\;$&$\xgap$&$\vol(\Oa)$&$\vol(\Og)$\\ 
\hline
 10 & /        & /         & 1.0      & /       & 0.0$^o$  &1.249046$^o$\\ 
 11 & -2.0     & -1.0      & 1.0      & 1.0     & 0.733333 &0.785398$^o$\\ 
 12 & -2.0     & -1.0      & 1.0      & 1.0     & 0.733333 &0.785398$^o$\\
 13 & -2.0     & -1.0      & 1.0      & 1.0     & 0.900000 &0.566729$^o$\\
 14 & -2.0     & -1.0      & 1.0      & 1.0     & 0.733333 &0.566729$^o$\\
 15 & -2.0     & -1.0      & 1.0      & 1.0     & 0.357143 & 0.554762   \\
 16 & -2.0     & -1.0      & 1.0      & 1.0     & 0.215320 & 0.334641   \\
\hline
\hline
\end{tabular} 
\lb{tab31c} 
\eeq

\noi From the tables \ref{tab31a}, \ref{tab31b}, \ref{tab31c} and Fig.1, 
we see that the volume of the cells tends to decrease montonically with 
increasing sequence length. The only exceptions are observed with length 
$N=13$, but are probably due to finite size effects. In two cases, a 
segment of a line of points in parameter space yields the same mxm-gap. 
\nl
In all the cases that we have considered, and where the ratios of the 
target potentials are rational numbers made up out of small enough 
integers, the maximization of the minimum gap renders the exact potentials.
Furthermore, we observe that all the obtained parameters are rational, 
even if the target parameters are not, due to the fact that in our
model only an integer number of contacts is possible. It also explains
the fact that for the target parameters $(-1,-1/\sqrt{2},0)$, the obtained 
parameter $E_{PP}$ is invariably $-5/7$ for all sequence lengths
$N=11,..,16$ although the cell changes drastically. One would have to 
consider (much) longer sequences to get contact numbers high enough to
generate a rational number closer to $-1/\sqrt{2}$.
This insensitivity may be lifted in cases where the number of contacts is 
no longer integer, e.g. for real space proteins.
\nl
For this set of target parameters, we have also considered taking only 
those good sequences with a minimum gap larger than certain threshholds 
(i.e. 0.5 and 0.75), and although the obtained cells are larger (it scales 
with $({\rm mingap})^{N_p}$), the obtained parameters are unaltered untill 
the cell ceases to be closed.
\nl\nl
To get an idea of the performance of the algorithm as the dimension of 
parameter space increases, we did some checks on the following variations:
\nl 
-a model with $N_p=3$, with 2 kinds of amino acids as before ($N_a=2$), but 
including a next to nearest neighbor (nnn) \nl
\hspace*{1mm} interaction for a H-P contact,\nl
-a model with $N_p=5$, with 2 kinds of amino acids ($N_a=2$) and nn and nnn 
contacts,\nl
-a model with $N_p=5$, with 3 kinds of amino acids ($N_a=3$) and only nn-contacts, 
and\nl 
- models\, with $N_p=9$, with 4 kinds of amino acids ($N_a=4$) and only 
nn-contacts, see Sec. 3.2\ .
\nl
The quality of the obtained parameters is always as good as those shown in 
tables \ref{tab31a},\ref{tab31b},\ref{tab31c} and does not 
depend on $N_p$. 
\nl
To check the sensitivity of the method to a wrong choice of energy 
function, we have generated good sequences and structures using 6 
interaction parameters ($N_p=5$, both $N_a=2$, nn- and nnn-contacts and 
$N_a=3$, nn-contacts), and tried to satisfy all inequalities using fewer 
parameters, e.g. ignoring nnn H-P contacts. The method immediately 
indicated that the cell does not exist, and thus that the number of 
parameters was insufficient. On the other hand, putting in more free 
parameters than were used to generate the good sequences, the irrelevance 
of these parameters is immediately recognized and their obtained values are
(very close to) 0\ .

\subsection{Results for the 4 amino acids problem}

The $P_{\mu\nu}$ matrix has 10 independent parameters in this case. 
Our tests have been carried out for four different sets of parameter
values where each parameter is generated independently from a Gaussian
distribution with mean $-2$ and variance $1$. The length of the chain is 14.
For each set of parameters, we have generated a PDB of about 600 sequences
and their corresponding (unique) native states. Furthermore, the
sequences have an energy gap, $\Delta$ (the energy difference between
the first excited state and the ground state) greater then 0.5. Indeed
it is thought \cite{SS} that real proteins in order to have
thermodynamical stability and short folding times should possess a
pronounced global minimum on the potential surface.
A comparison with one case where $\Delta > 2$ is also presented.
The trial Hamiltonian is parametrized as the true one and we have chosen
the energy scale by fixing to its exact value one of the most negative
$P_{\mu\nu}$'s.
The remaining $9$ parameters are then determined maximizing the minimum
gap using the method explained above.
We have also verified that simulated annealing techniques are quite
efficient for this case and give the same set of extracted potentials as
the method used in \S \ref{reshp}.
Figures 2a,b,c and d show the extracted potentials versus the true ones.
The extracted potentials are then tested for {\it new sets} of "good"
sequences for each of the four cases to determine their ground state
configurations over all possible self avoiding chains of length 14.
For all the sequences in the PDB, we get full success (Figure 2). 
Indeed,
since the maximun gap has been calculated on a restricted set of
conformations there is no guarantee a priori that the good sequences
used in the optimization procedure recognize their own native state
among all possible conformations. Thus it is important to test the
extracted potentials using a new independent set of good sequences.
In all four cases that we studied,  at most $2$ out of $628$ do not found their original native state. The percentages of the correct determination of the native
states using the extracted potentials are indicated in the table.

\renewcommand{\theequation}{3.2}

\beq
\begin{tabular}
{||l | l       | l    ||}
\hline
\hline
{\rm parameter\ set}&{\rm size\ of\ the\ PDB}&{\rm success} \\
\hline
 1 & 628 & 99.7\% \\ 
 2 & 716 & 99.9\% \\
 3 & 840 & 100\%  \\
 4 & 798 & 100\%  \\
\hline
\hline
\end{tabular} 
\lb{tab32}
\eeq

\renewcommand{\theequation}{\arabic{equation}}
\setcounter{equation}{10} 

\noi We have tested the performance of the method as the size of the PDB is
decreased. Figure 3 shows how the percentage of success depends on the
number $N$ of sequences contained in the PDB.  Only three of the four
cases are shown for clarity (the fourth case has the same behaviour as
the other three). The minimum $N$ used is $14$. Note that full success is 
almost reached for $N\sim 200-300$. For the first set of potential 
parameters, we have also generated a PDB with an energy gap $\Delta>2$. 
The results are shown in fig.3\ , and saturation is reached at about 
$N\sim 100$.

\section{Comparison with other methods                           \lb{comp}}

\subsection{The quasi-chemical method}

The quasi-chemical method [5-7] is widely used in various forms for 
obtaining the effective potential between aminoacids and to provide
``scores'' for candidate protein structures. Briefly, the procedure is as 
follows: from the databank, one compiles the probability density,
$f_{A,B}(r),$ that two specific aminoacids are at a distance $r$ from each
other. This quantity is a normalized one, and takes into account how often 
the individual aminoacids appear in the data base. The basic idea is that 
if $A$ and $B$ like each other, they are more likely to be near each other 
compared to a random reference state of a non-interacting gas of aminoacids.
Conversely, if $A$ and $B$ dislike each other, they avoid each other a bit 
more than what one would expect from random considerations. This idea is 
then quantified in the form 

\beq
E_{A,B}(r)\propto -kT\ln[f_{A,B}(r)]\ . \lb{q}
\eeq

\noi Additional considerations pertaining to how far apart two aminoacids 
are along the sequence are sometimes introduced in order to build in the 
correct secondary structure. The derived quantities such as $E_{A,B}$ are 
now interpreted as the energies of interaction between the aminoacid pairs 
and used for determining which structure among many alternatives yields the 
most favorable value of the energy. The results using this method as 
obtained by Thomas and Dill \cite{TD} are shown in table \ref{tab41}, and 
have to be compared with the corresponding cases in tables \ref{tab31a},
\ref{tab31b},\ref{tab31c} obtained by our 
method.

\renewcommand{\theequation}{4.1}

\beq
\begin{tabular}
{|| c    | c        |c         ||c        |c         |c         |c       ||}
\hline\hline
\multicolumn{3}{||c||}{\rm True}&\multicolumn{4}{|c||}{\rm TD\ test
\cite{TD}}                                                                \\
\hline\hline
$E_{HH}$ & $E_{HP}$ & $E_{PP}$ & $E_{HH}$ & $E_{HP}$ &$E_{PP}$  & success \\
\hline
  -5     &   -4     &   -1     &   -5     &   -3.0   &   +0.8   &   74\%  \\
  -5     &   -1     &   -2     &   -5     &   -1.1   &   -2.1   &  100\%  \\
  -5     &   -5     &   -1     &   -5     &   -3.7   &   +1.4   &   84\%  \\
  -5     &   -3     &   +1     &   -5     &   -2.6   &   +2.5   &   96\%  \\
  -5     &   -3     &   -1     &   -5     &   -2.4   &    0.0   &   64\%  \\
\hline\hline
\end{tabular}
\lb{tab41}
\eeq

\renewcommand{\theequation}{\arabic{equation}}
\setcounter{equation}{11} 

\noi The rationalization for deriving the interaction energy from the 
observed pairing frequency has been stated to be Boltzmann's principle 
and also has been called the Boltzmann device.  Boltzmann statistics 
pertains to the occupation probabilities for the energy levels of an 
individual system. Thus, if a system can have energies $E_0$, $E_1$, $E_2$,
etc., the probability that the system has an energy $E_i$ is proportional 
to $\exp[-{E_i\over kT}]$.  
\nl\nl
We repeat several observations made in Seno et al. \cite{SM}. (Thomas and 
Dill \cite{TD} have also presented an important critique of the 
quasichemical method.) First, the native 
structures of distinct sequences of aminoacids do not correspond to the 
excitations of a single system. Instead, each of the sequences is a 
separate system, whose native state structure is known from experiment.
Thus, the basic premise of the method is wrong. Second, even making 
the assumption that Boltzmann statistics did hold, there is no simple 
relationship between the observed pairing frequency and the energy of 
interaction as envisaged by (\ref{q}). 
The role of temperature in (\ref{q}) is unclear, because the 
native states of each of the sequences correspond to their ground states or
equilibrium states at $T=0$. Third, the quasichemical method relies on a 
reference state -- the observed pairing frequencies are compared to those 
expected in this reference state in order to determine whether two amino
acids like each other and by how much.  Often, this reference state is 
chosen as a noninteracting gas made up of all the aminoacids constituting 
all the sequences with known native structure. This does not seem to have a
physical basis, because the sequences are all distinct entities and do not 
originate from a common soup of aminoacids.
\nl\nl
These difficulties with the quasichemical method, which were already 
partially recognized in the literature (see ref. \cite{TD} and references 
therein), are avoided in our strategy. The sequences whose structures are 
known are analogous to quenched variables in statistical mechanics while 
the conformations that a given sequence can adopt, are the analog of 
annealed variables. A thermodynamic average can be performed over the 
annealed variables but not over the quenched ones. We use Boltzmann 
statistics but for each sequence separately. We deal with the energies 
directly and not with a derived quantity such as the pairing frequency. 
Indeed, our strategy embodies the complete information in the system and,
in principle, has information not only about pairing frequencies but also 
triplet and higher order correlations. Our method does not rely on a 
reference state and the role of temperature is well-defined. 

\subsection{Mirny and Shakhnovich's method}

Recently, Mirny and Shakhnovich (MS) \cite{MS} have proposed to use the 
$Z$-score \cite{BL}, which is a measure of how pronounced the energy 
minimum corresponding to the native state is, to carry out protein design.
The $Z$-score is given by:

\beq
Z(\bsi)={E(\Ga^n_\bsi,\bsi)-\Eh\over{\rm var}(E)}\ .
\lb{Z} 
\eeq

\noi where the average of $E$, $\Eh$, and its variance, var$(E)$, are 
computed for a set of alternative (decoy) conformations. The method of 
\cite{MS} entails the minimization of the cost function (harmonic mean)

\beq
\Zh= (\sum_{\bsi\in \Om_\s} Z(\bsi)^{-1})^{-1}\ .
\lb{Zh}
\eeq

\noi For each conformation $\Ga$ of the ensemble $\Om_\s$, the average 
$\Eh$ and the variance var$(E)$ are calculated in an ensemble of phantom 
conformations with the same number of residue-residue contacts as in $\Ga$ 
and with the assumption that these contacts occur independently of each 
other. This approximation will be discussed further later on. One can 
repeat these calculations for our cases.
\nl
We have implemented this method using (\ref{Z}) and (\ref{Zh}), calculating
$\Eh$ and var$(E)$ exactly for each sequence using the structures of the 
PDB. 
\nl
In order to have a finite minimum of $\Zh$, it is necessary to fix the 
variance, or equivalently one of the interaction parameters like we did. 
MS also fix the average potential, which requires more information, which, 
apriori, one does not have. In the case of many interaction parameters, 
however, this might not be so crucial. Furthermore, in their implementation,
MS do not explicitly require that all the $Z(\bsi)$'s be negative, since 
$\Eh$ and var$(E)$ are approximated. For the H-P model, we first require 
that $Z(\bsi)<0$ for all $\bsi$ in the PDB, and then we minimize $\Zh$. 

\renewcommand{\theequation}{4.2.\alph{equation}}
\setcounter{equation}{0} 

\beq
\begin{tabular}
{||l | l      | l       | l      | r      | r        | l         ||}
\hline
\hline
{\rm true}&-1.0&-0.707107& 0.0   &        &          &           \\ 
\hline
 $N$& $E_{HH}$& $E_{PP}$&$E_{HP}$&$\#_{tot}$&$\#_{wrong}$&success\\
\hline
 12 &-0.969868&-0.747827&0.010588&  728   &   4      & 99.45\%   \\
 13 &-0.990930&-0.719754&0.003577&  750   &   0      & 100.0\%   \\
 14 &-0.977091&-0.738392&0.008377& 2005   &  26      & 98.70\%   \\
 15 &-0.963963&-0.755398&0.012254& 4302   &  77      & 98.21\%   \\
 16 &-0.972729&-0.744113&0.009735& 8892   & 151      & 98.30\%   \\
\hline
\hline
\end{tabular}
\lb{tab42a}
\eeq

\beq
\begin{tabular}
{||l | l      | l       | l      | r      | r        | l         ||}
\hline
\hline
{\rm true}&-1.0&-0.707107& 0.0   &        &          &           \\ 
\hline
$N$ & $E_{HH}$& $E_{PP}$&$E_{HP}$&$\#_{tot}$&$\#_{wrong}$& success\\
\hline
 12 &-0.655207&-0.330015&0.474504&  728   &  124     & 82.97\%   \\
 13 &-0.841471&-0.317697&0.568875&  750   &  198     & 73.60\%   \\
 14 &-0.745093&-0.317782&0.526300& 2005   &  675     & 66.33\%   \\
 15 &-0.657765&-0.327509&0.477554& 4302   & 1754     & 59.23\%   \\
 16 &-0.740636&-0.328271&0.517735& 8892   & 3274     & 64.00\%   \\
\hline
\hline
\end{tabular}
\lb{tab42b}
\eeq

\renewcommand{\theequation}{\arabic{equation}}
\setcounter{equation}{13} 

\noi For the 4 amino acid case, the minimization of (\ref{Zh}) leads to 
spurious minima corresponding to $\sum_\bsi Z(\bsi)^{-1}\simeq0$, since 
both positive and negative $Z(\bsi)$'s appear. This happens both with the 
exact $\Eh$ and var$(E)$\ , and with the approximations of MS \cite{MS}.
\nl
Since the search of the parameter domain where all the $Z(\bsi)$'s are 
negative was impractical in this case, we have modified (\ref{Zh}) to
$\langle|Z|\rangle_{\rm harm}=(\sum_\bsi|Z(\bsi)|^{-1})^{-1}$\ .
\nl
Tables \ref{tab42a},\ref{tab42b} show the results for one of the H-P cases we have considered 
before, i.e. $E_{HH}=-1,\ E_{HP}=-1/\sqrt{2}$ and $E_{PP}=0$\ . 
Table \ref{tab42a} corresponds to the case where we have fixed both the variance 
and the average of the $E$'s, leaving only one parameter to be determined. 
Table \ref{tab42b} shows the results when only the variance of the interaction 
parameters is fixed to set the energy scale, and two parameters are left to
be determined, as in our method.
\nl
With the parameters obtained from the minimization we checked how many of 
the good sequences of the PDB still have their unique ground state in the 
correct conformation among all the possible conformations obtained by exact
enumeration. In contrast to our method, not all of the sequences in the PDB
find the correct native state, as can be seen in the tables. Note that neither 
the success rate, nor the values of the potentials are monotonic 
as a function of
the chain length, at least within the small range of lengths used.
\nl\nl
Table \ref{tab42c} shows the results for the $4$ amino acid case for the same
four sets of potential parameters used to test our method. The variance
and the average potential have been fixed to their exact values (thus
there are $8$ free parameters and not $9$ as in our case). For one
parameter set we have also implemented the MS optimization method. MS
use the following expressions for $\Eh$ and var$(E)$
(see the discussion following eq.\ref{Zh} ):

\bea
\Eh&=&\sum_{i<j} P_{\mu_i,\mu_j} \langle \Delta_{i,j}\rangle \\ 
{\rm var}(E)&=&\sum_{i<j}\sum_{k<l}P_{\mu_{i},\mu_j}P_{\mu_k,\mu_l}T_{ij,kl}
\eea

\noi with 

\beq
\langle\Delta_{i,j}\rangle={n\over n_{tot}}
\lb{deltaMS}
\eeq

\noi and

\beq
T_{ij,kl} = \left\{ 
\begin{array}{ll}
\frac{1}{n_{tot}^{2}}\hsp{27.5mm}(i,j)\neq(k,l) \\
\\
\frac{1}{n_{tot}}-\frac{1}{n_{tot}^2}\hsp{16mm}(i,j)=(k,l)
\end{array}\right. 
\lb{tMSt}
\eeq

\noi where $n$ is the number of contacts in the native conformation,
$n_{tot}$ is the total number of the topologically possible contacts and
the indices $i,j,\dots$ run from $1$ to the length of the chain ($14$ in 
our case). Using the MS hypothesis, we found a different expression for 
$T_{ij,kl}$:

\beq
T_{ij,kl} = \left\{ 
\begin{array}{ll}
{n(n-1)\over n_{tot}(n_{tot}-1)}-{n^2\over n_{tot}^2}\hsp{5.5mm}(i,j)\neq
(k,l)\\
\\
{n\over n_{tot}}-{n^2\over n_{tot}^2}\hsp{17mm}(i,j)=(k,l)
\end{array}\right. 
\lb{tMSw}
\eeq

\noi The results corresponding to both assignments, (\ref{tMSw}) and 
(\ref{tMSt}), are also reported in table \ref{tab42c} and should be compared
with the results of our method in table \ref{tab32}\ .
Figures 4. a, b, c and d are the analogs of fig. 2. a, b, c and d for the MS
method. Figure 4.a shows the extracted potentials using both the exact
$\Eh$ and var$(E)$ and the approximation (\ref{deltaMS}) and (\ref{tMSw}) 
(which according to table \ref{tab42c} works better than the MS one, i.e. 
(\ref{deltaMS}) and (\ref{tMSt})) for parameter set $1$.

\renewcommand{\theequation}{4.2.\alph{equation}}
\setcounter{equation}{2} 

\beq
\begin{tabular}
{||l | l       | l         | l  ||}
\hline
\hline
parameter set &$\#_{tot}$&$\#_{wrong}$&success\\
\hline
 1 & 628 & 64 & 89.8 \%\\ 
 2 & 716 & 80 & 88.2 \%\\
 3 & 840 & 14 & 98   \%\\
 4 & 798 & 96 & 88   \%\\
\hline
 1 (using eq \ref{tMSw}) & 628 & 71  & 88 \%\\  
\hline
 1 (using eq \ref{tMSt}) & 628 & 105 & 83 \%\\
\hline
\hline
\end{tabular} 
\lb{tab42c}
\eeq


\section{Acknowledgements}
One of us (C.C.)  warmly acknowledges
Giovanni Fossati for advice and suggestions.
This work was supported by grants from INFM, NASA, NATO,
The Petroleum Research Fund administered by the American Chemical Society, The Office of Naval Research
through the Applied Research Laboratory at Penn State and the Center for
Academic Computing at Penn State.

\newpage
\begin{center}
figure captions
\end{center}

\noi{\footnotesize
{\bf Fig.1} \hsh
The 2 dimensional cells for the target parameters ($E_{HH}=-1,E_{PP}=0,
E_{HP}=0$) (in units of $k_B T$), using all the structures as alternatives, 
for different sequence lengths. Indicated are the cell (shaded area), the 
target parameters (fat point), the sequence length (10, 12, 14, 16), the 
volume (if the cell is closed) or the opening angle alpha of the allowed 
area. 
}
\nl\nl

\noi{\footnotesize
{\bf Fig.2} \hsh
Derived potential versus true potential for the 4 aminoacid problem (in 
units of $k_B T$). Results for the parameter set 1 (a), 2 (b), 3 (c) and 4 
(d).
}
\nl\nl

\noi{\footnotesize
{\bf Fig.3} \hsh
Effect of the PDB size on the success rate using the extracted parameters 
to determine the ground state configurations for new sets of sequences, 
for different minimum energy gaps $\Delta$ (in units of $k_B T$). For the 
case with $\Delta>2$ (open triangles), the cell is not closed for small 
$N$. When the cell is closed, the success rate is almost 100
}
\nl\nl

\noi{\footnotesize
{\bf Fig.4} \hsh
Derived potential versus true potential (in units of $k_B T$), for the 4 
aminoacid problem using the MS method. Results for the parameter set 1 (a), 
2 (b), 3 (c) and 4 (d). Fig. 1.a also shows the results using the 
approximation (\ref{deltaMS}) and (\ref{tMSw}) (open circles).
}
\nl\nl

\begin{center}
table captions
\end{center}

\noi{\footnotesize
{\bf Tab. \ref{tab31a},\ref{tab31b},\ref{tab31c}} \hsh 
Results for the H-P model fixing $E_{HH}$ to its true value to set the 
energy scale (in units of $k_B T$). The table shows the sequence length, 
the obtained interaction parameters, the true minimum gap with the target 
parameters, the obtained mxm-gap, the volume of the cell (or opening angle 
in cases in which the cell is not closed) both using all conformations and 
only the ``good'' ones as alternative conformations. The success rate in 
the prediction of native conformations of the ``good'' sequences with the 
obtained parameters is 100\% in all cases that the cell is closed.
}
\nl\nl

\noi{\footnotesize
{\bf Tab. \ref{tab32}} \hsh 
Results for the 4 aminoacid model.  For each parameter set, the table
shows the size of the PDB used for the derivation of the potential and the 
success rate in the correct prediction of the native state for each of the 
training set sequences.
}
\nl\nl

\noi{\footnotesize
{\bf Tab. \ref{tab41}} \hsh
Summary of the results of TD \cite{TD} using the 
Miyazawa-Jernigan scheme \cite{MJ}, for a sequence length of 14 monomers. 
The table shows the true parameters, the parameters obtained by
TD and the success rate in the prediction of native conformations of the 
``good'' sequences with the obtained parameters.
}
\nl\nl

\noi{\footnotesize
{\bf Tab. \ref{tab42a}} \hsh
Results for the H-P model, using the $Z$-score of MS \cite{MS}, fixing both the 
variance and the average of the interaction parameters to their true values.
The table displays the sequence length, the derived interaction 
parameters, the total number of good sequences, the number of sequences
for which the predicted ground state is wrong, and the success rate.
}
\nl\nl

\noi{\footnotesize
{\bf Tab. \ref{tab42b}} \hsh 
Same as Tab. \ref{tab42a}, but only fixing the variance of the interaction 
parameters.
}
\nl\nl

\noi{\footnotesize
{\bf Tab. \ref{tab42c}} \hsh 
Results for the 4 aminoacid model, using the MS method \cite{MS}, 
fixing both the variance and the average of the interaction parameters to 
their true values. The table shows  the identification of the parameter 
set, the size of the PDB used for the derivation of the potentials, the 
number of failures in the prediction of the correct native state,
and the success rate.
In the top 4 lines, the exact $\Eh$ and var$(E)$ are used, whereas  
for the fifth line the approximation (\ref{deltaMS}) and (\ref{tMSw}), and
for the sixth line the approximation (\ref{deltaMS}) and (\ref{tMSt}) are
used.
}
\newpage .

\newpage
\begin{thebibliography}{99}

\bibitem{An} 
C. Anfinsen, Science {\bf 181}, 223(1973).
\bibitem{SS}
A. Sali, E. Shakhnovich and M. Karplus, Nature {\bf 369}, 248 (1994);
J. Mol. Biol. {\bf 235}, 1614 (1994).
\bibitem{DB}
K.A. Dill, S. Bromberg, S. Yue, K. Fiebig, K.M. Yee, P.D. Thomas and H.S.
Chan, Protein Science {\bf 4}, 561 (1995).
\bibitem{GO}
T. Garel, H. Orland and D. Thirumalai, New Developments in Theoretical 
Studies of Proteins, R. Elber (ed.), World Scientific, Singapore, (1996). 
\bibitem{TS} 
S. Tanaka and  H.A. Scheraga, Macromolecules {\bf 9}, 945 (1976).
\bibitem{MJ}
S. Miyazawa and R.L. Jernigan , Macromolecules {\bf 18}, 
534 (1985); J. Mol. Biol. {\bf 256}, 623 (1996).
\bibitem{XX} 
M.J. Sippl, J. Mol. Biol. {\bf 213},  859 (1990);
M. Hendlick,  P. Lackner, S.  Weitckus , H.  Floeckner,
R. Froschauerer, K.   Gottsbacher,  G. Casari and  M.J. Sipp,
J. Mol. Biol {\bf 216},  167 (1990); 
D.A. Hinds and M. Levitt, Proc.
Natl. Acad. Sci. USA {\bf 89}, 2536 (1992);
D.T.  Jones,  W.R. Taylor and  J.M. Thorton, Nature
{\bf 358}, 86 (1992);
A. Godzik and  J. Skolnick , Proc. Natl. Acad. Sci. USA {\bf 89},
12098 (1992);
S.H. Bryant and  C.E. Lawrence, Proteins: Struct. Funct. Genet. {\bf 16}, 
92 (1993);
M. Wilmanns and  D. Eisenberg, Proc. Natl. Acad. Sci. USA {\bf 90}, 
1379 (1993);
M. Pellegrini , S.  Doniach, Proteins: Struct. Funct. Genet. {\bf 15},
436 (1993);
K. Nishikawa and Y. Matsuo, Protein Eng. {\bf 6},811  (1993);
M.J. Sippl, J. Comput. Aided Mol. Des. {\bf 7}, 
473 (1993); Proteins {\bf 17}, 355 (1993);
M.J. Sippl, M. Jaritz, in {\em Protein structure
by distance analysis}. Edited by H. Bohr and S. Brunak , (1994) Amsterdam
IOS press;
M.J.  Sippl, M. Jaritz , M. Hendlich, M. Ortner, P. Lackner 
in {\it Statistical Mechanics, protein structure and protein
substrate interactions} Edited by S. Doniach, (1994), New York: Plenum
Publishers;
M.J. Sippl , Current Opinion in Structural Biology {\bf 5},
229 (1995) and references therein;
H.  Flockner, M. Braxenthaler, P. Lackner, M. Jaritz,
M. Ortner and M.J. Sippl, Proteins: Structure, Function and Genetics
{\bf 23}, 376 (1995);
P.D. Thomas and  K. Dill, An Iterative Method
for Extracting Energy-like Quantities
from Protein Structures, Preprint (1996).
\bibitem{XY} 
M.S. Friederichs and P.G. Wolynes, Science {\bf 246}, 371 (1989);
R.A. Goldstein, Z.A. Luthey-Schulten and  P.G. Wolynes,
 Proc. Nat. Acad. Sci.
U.S.A  {\bf 89},  4918 (1992) have used spin glass theory to maximize the 
stability gap of the native protein structures with respect to average
decoy structures;
R.A. Goldstein, Z.A. Luthey-Schulten and P.G. Wolynes,  Proc.
Natl. Acad. Sci. USA {\bf 89}, 9029 (1992).
\bibitem{Cr}
G.M. Crippen, Biochemistry {\bf 30}, 4232 (1991).
\bibitem{MC}
V.N. Maiorov and G.M. Crippen, J. Mol. Biol. {\bf 227}, 876 (1992).
\bibitem{C2}
G.M. Crippen, Protein: Struct. Funct. Genet. {\bf 26}, 167 (1996).
\bibitem{DK}
J. M. Deutsch and T. Kurosky, Phys. Rev. Lett. {\bf 76}, 323 (1996);
F. Seno, M. Vendruscolo, A. Maritan and J.R. Banavar, Phys. Rev. Lett.
{\bf 77}, 1901 (1996); M.P. Morrisey and E.I. Shakhnovich, Folding and
Design {\bf 1}, 391 (1996).
\bibitem{SR} 
R. Srinivasan and G.D. Rose, Proteins: Structures, Function and 
Genetics {\bf 22}, 81 (1995).
\bibitem{TD} 
P.D. Thomas  and  K.  Dill, J. Mol. Biol. {\bf 257}, 457  (1996).
\bibitem{MS} 
L.A. Mirny  and E.I. Shakhnovich , 
J. Mol. Biol. {\bf 264}, 1164 (1996).
\bibitem{SM} F. Seno, A. Maritan and J.R. Banavar, Proteins: Structure, 
Function, and Genetics, in press.
\bibitem{LD}  
K.F. Lau and K.A. Dill, Macromolecules {\bf 22},
3986 (1989).
\bibitem{BL} J.U. Bowie, R. Luthy and D. Eisenberg, Science {\bf 253},
164 (1991).

\end{thebibliography}
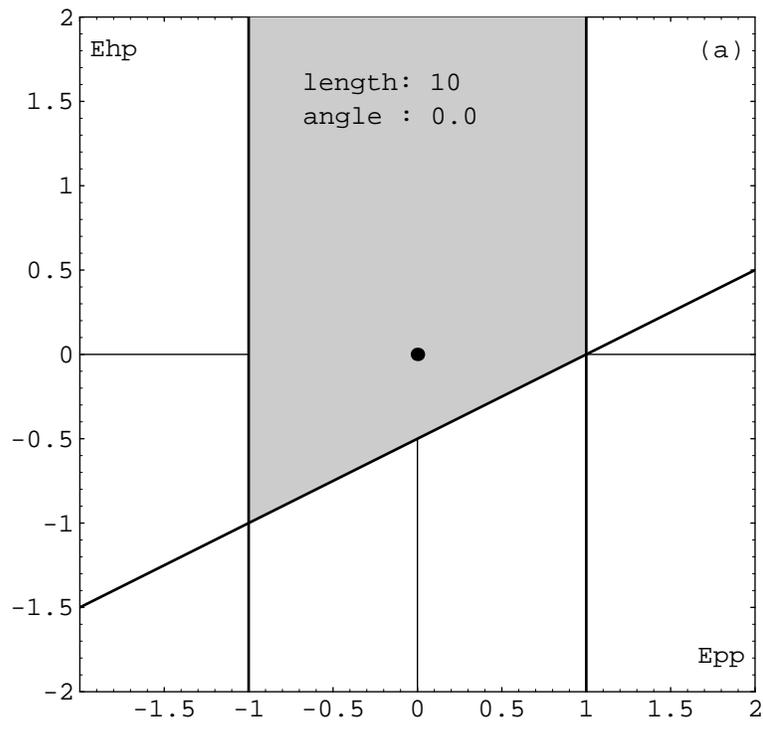
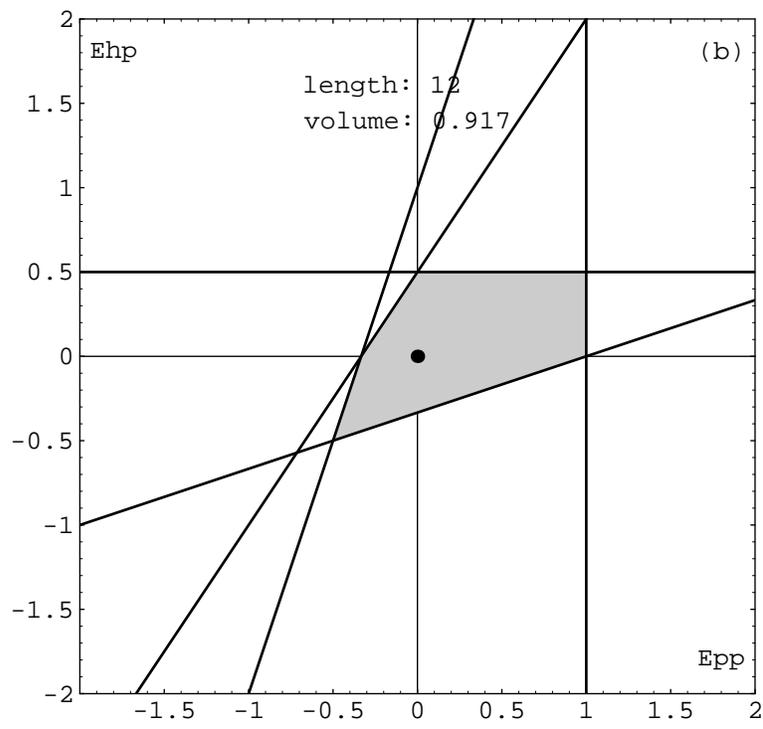
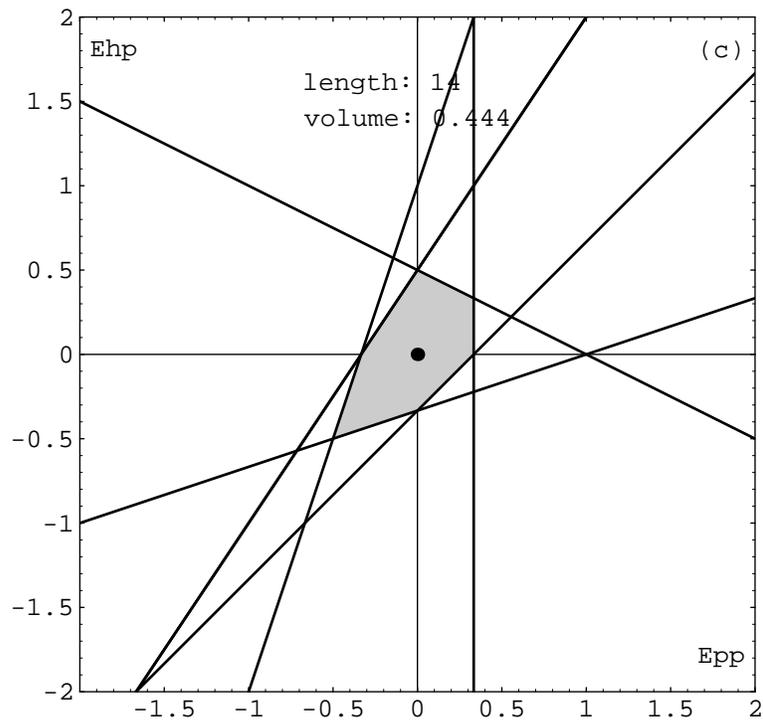
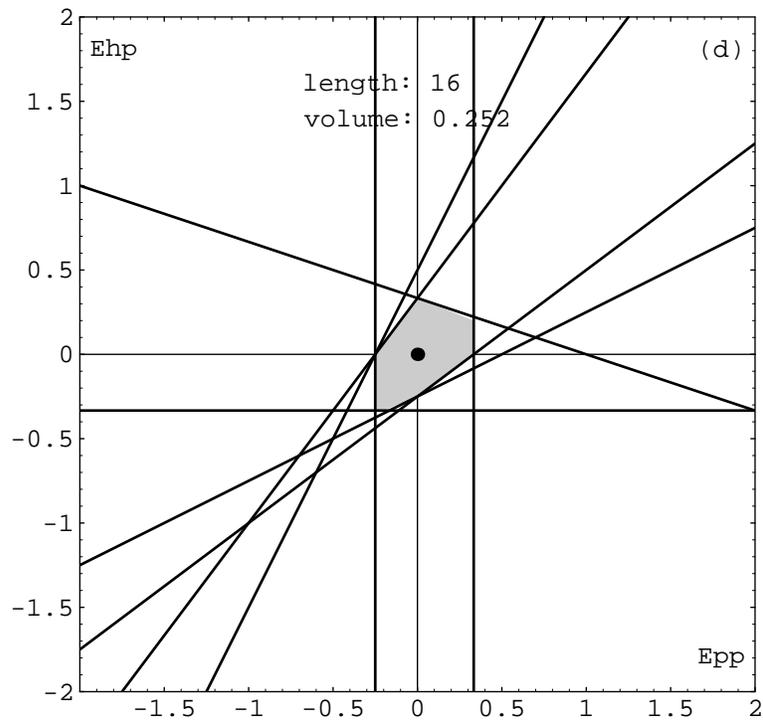
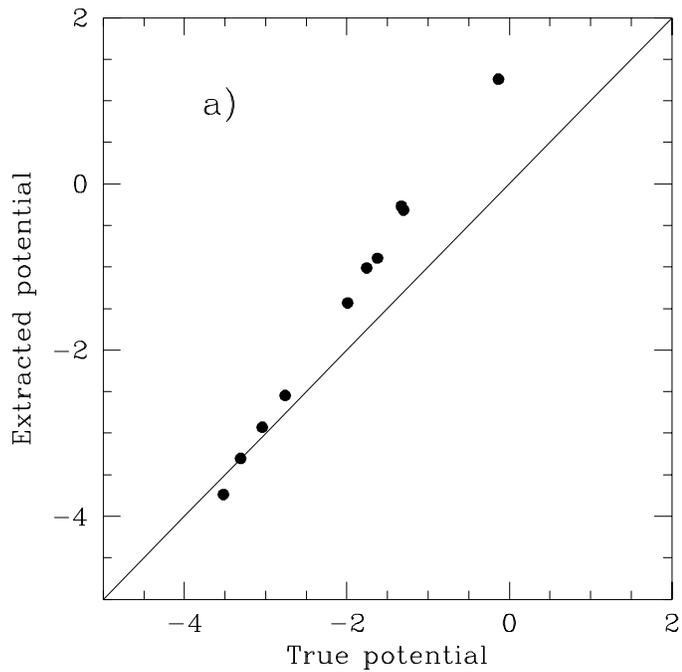
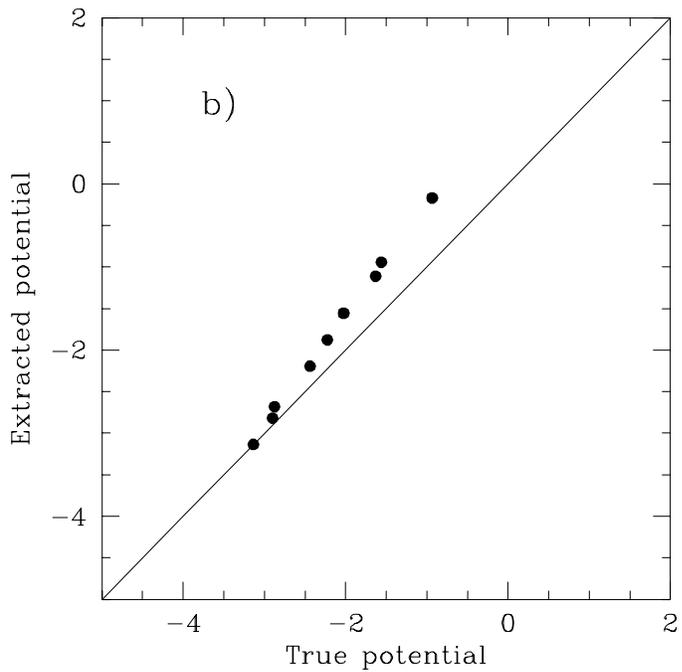
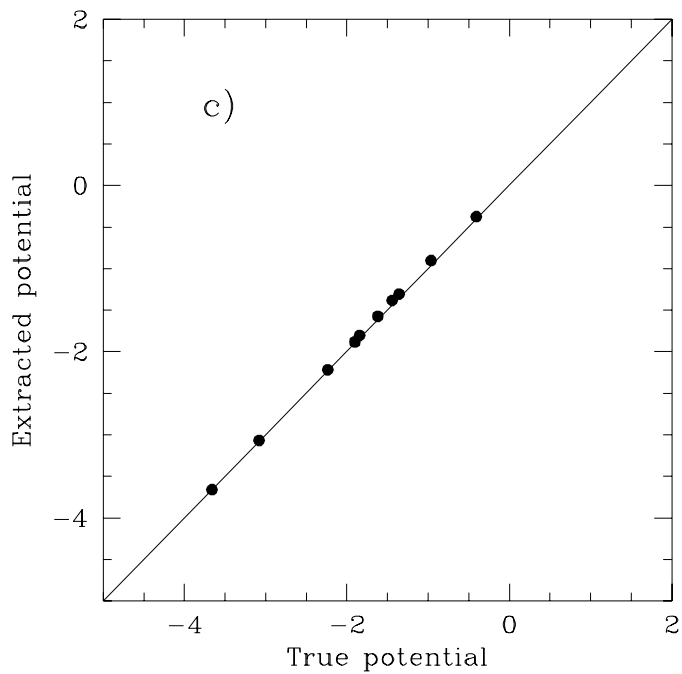
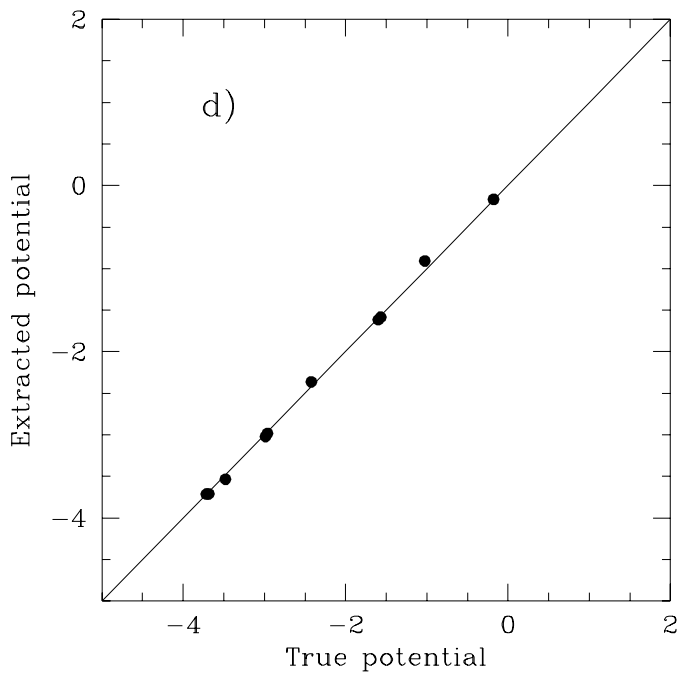
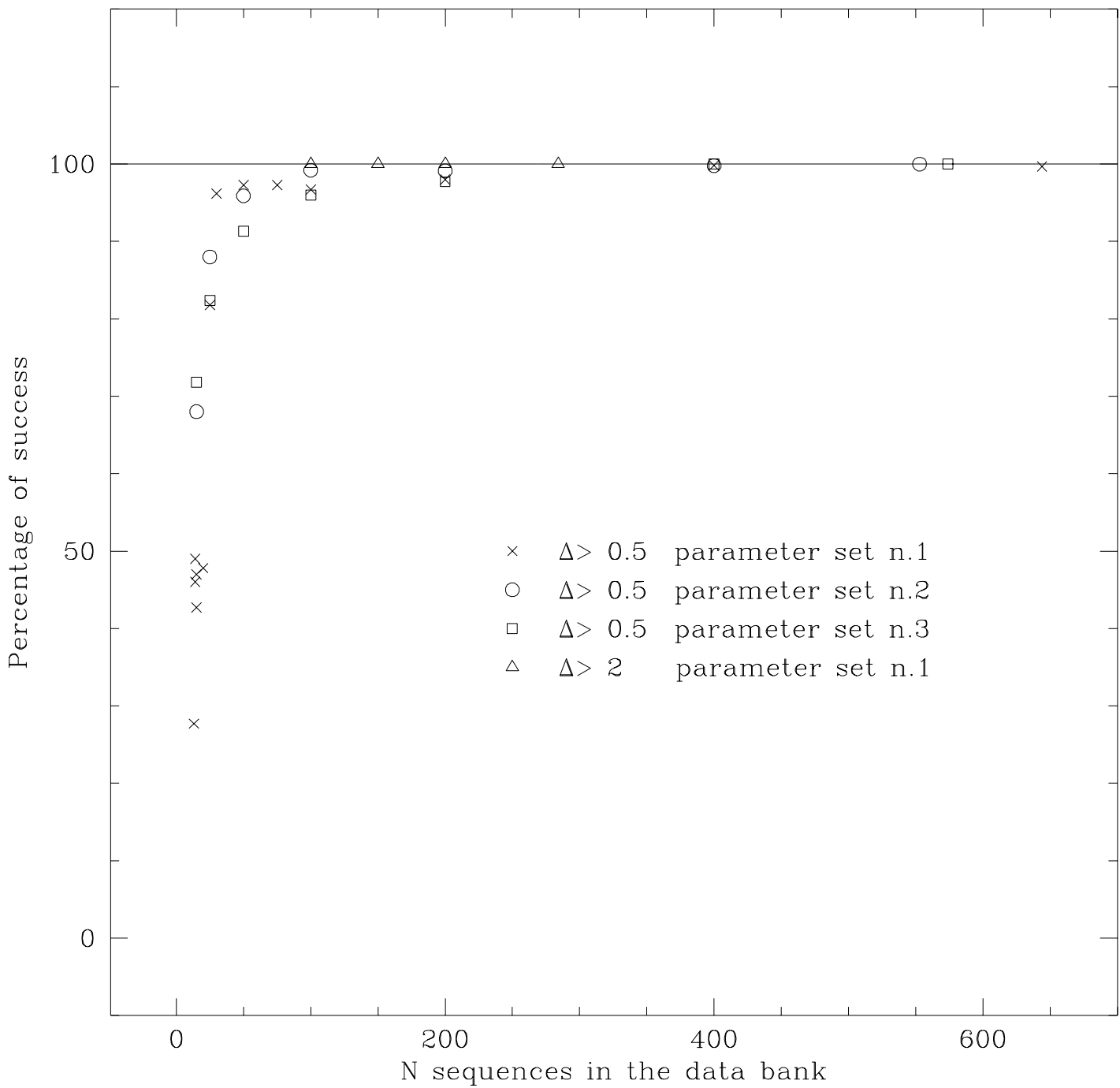
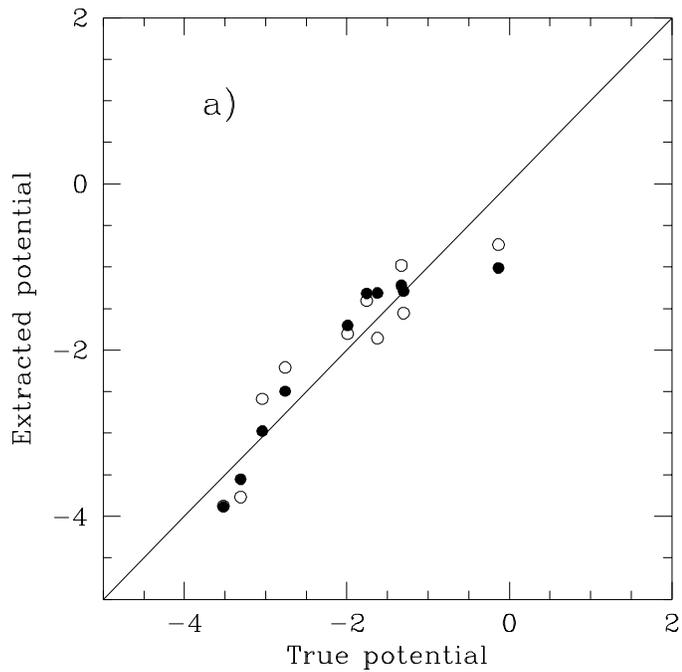
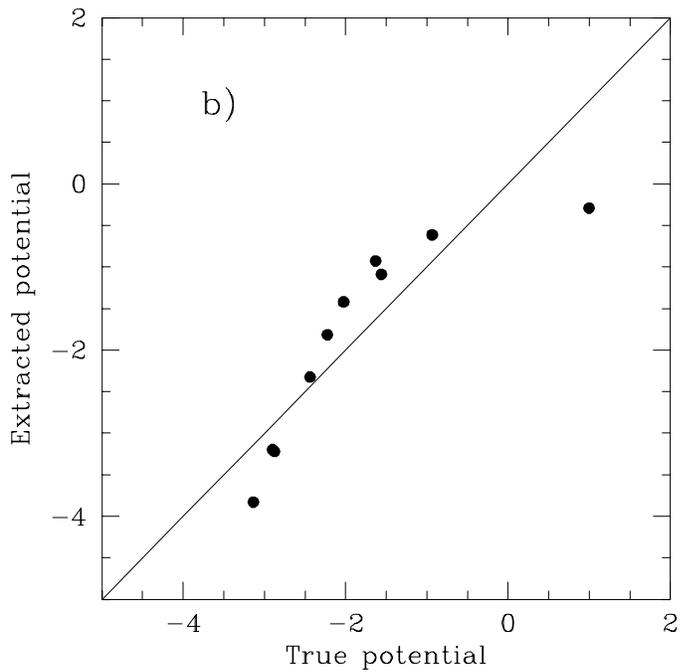
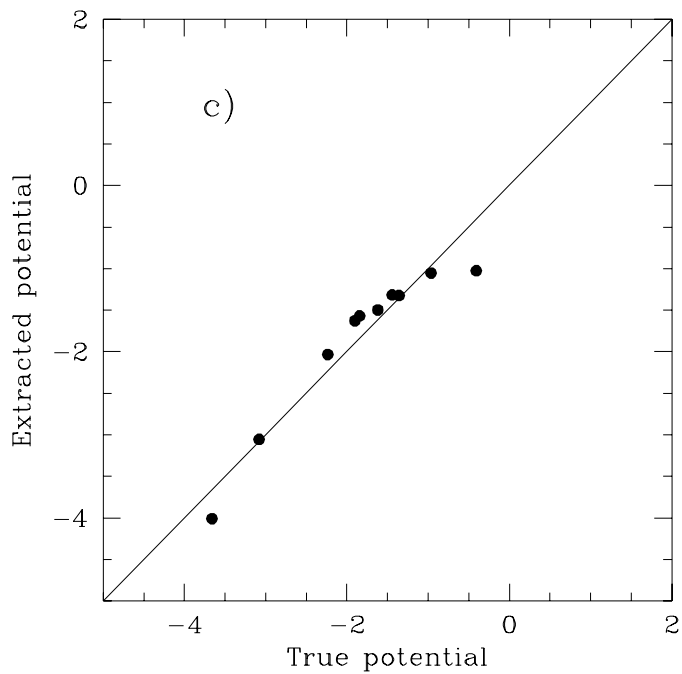
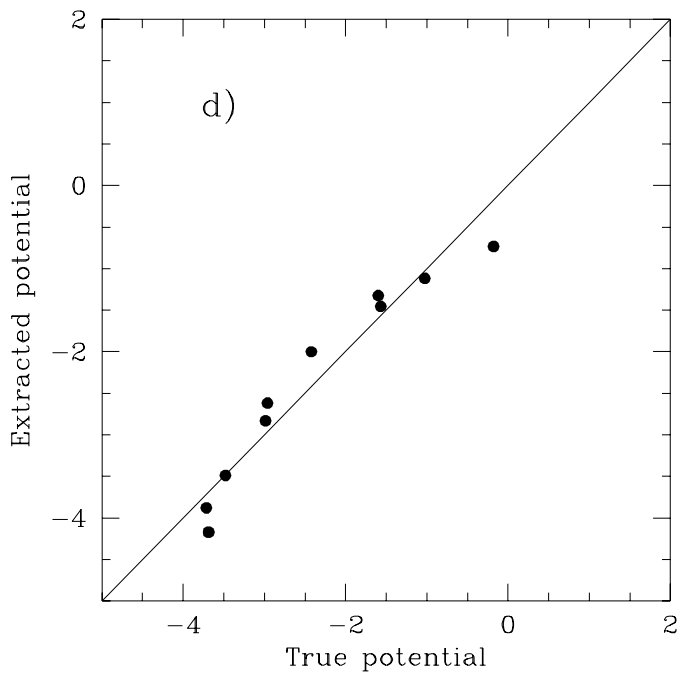
\end{document}